\documentclass[11pt]{article}
\usepackage{cospar}
\usepackage{url}
\usepackage{txfonts}
\usepackage[sectionbib]{natbib}
\usepackage{graphicx}

\title{FINE STRUCTURE OF THE THERMAL AND NON-THERMAL X-RAYS
IN THE SN~1006 SHELL}

\author{A. Bamba\address{Department of Physics, Kyoto University,
	Kitashirakawa, Sakyo-ku, Kyoto, 606-8502, Japan},
        R. Yamazaki$^{1}$, M. Ueno$^{1}$, and K. Koyama$^{1}$}

\begin{document}

\maketitle

\begin{abstract}
The North Eastern shell of SN~1006 is the most probable acceleration site
of high energy electrons (up to $\sim$ 100~TeV)
with the Fermi acceleration mechanism at the shock front.
We resolved non-thermal filaments from thermal emission in the shell
with the excellent spatial resolution of $Chandra$.
The non-thermal filaments seem  thin sheets with the scales of
$\sim 1$~arcsec (0.01~pc) and $\sim 20$~arcsec (0.19~pc)
upstream and downstream of the shock, respectively.   
In a simple diffusive shock acceleration (DSA) model
with the magnetic field parallel to the shock normal,
the downstream region should have a highly disordered magnetic field
of 30--40~$\mu$G. 
The width at the upstream side is extremely small,
comparable to the gyro-radius of the maximum energy electrons.
This result might imply that
the possible scenario is not
the conventional diffusive shock acceleration with a parallel magnetic field
but that with a nearly perpendicular field
or electron shock surfing acceleration.
\end{abstract}

\section*{INTRODUCTION}
Since the discovery of cosmic rays,
the origin and acceleration mechanism up to energies of $10^{15.5}$~eV
(the ``knee'' energy) have been long-standing problems.
A breakthrough came from the X-ray studies of  SN~1006;
Koyama et al. (1995) discovered synchrotron X-rays from the shells of this
supernova remnant (SNR),
indicating the existence of extremely high energy electrons
up to TeV energies or more
produced by the first order Fermi acceleration.

The mechanism for cosmic ray acceleration has also been studied
for a long time
and the most plausible process is a diffusive shock acceleration (DSA)
(e.g.\ Bell, 1978).
Apart from the globally successful picture of DSA,
detailed but important processes,
such as the injection and the reflection of accelerated particles,
are not well understood.
The spatial distribution of accelerated particles
responsible for the non-thermal X-rays,
may provide key information on these unclear processes.
Previous observations, however, are too limited in spatial resolution
for a detailed study on the structure of shock acceleration process
and injection efficiency.
Although many observations and theoretical models are made for SN~1006,
these problems are still unsolved
(e.g. Reynolds, 1998).

In this paper, we report on the first results
of the spectral and spatial studies of
the non-thermal shock structure in the North Eastern (NE) shell of SN~1006
with  $Chandra$.
We discuss the spectral analyses and determine the scale lengths of the 
structures for non-thermal electrons
on the basis of a simple and a conventional DSA model.
In this paper, we assume the distance of SN~1006 to be 2.18~kpc
(Winkler et al. 2003).

\section*{OBSERVATION}
We used the $Chandra$ archival data of the ACIS
on the NE shell of SN~1006 (Observation ID = 00732)
observed on July 10--11, 2000 with the targeted position at
(RA, DEC) =
($15^{\rm h}03^{\rm m}51^{\rm s}.6$, $-41^{\rm d}51^{\rm m}18^{\rm s}.8$).
The satellite and instrument are described
by Weisskopf et al. (1996).   
CCD chips I2, I3, S1, S2, S3, and S4 were used with the pointing center on S3.
Data acquisition from the ACIS was made in the Timed-Exposure Faint mode
with the readout time of 3.24~s.
The data reductions and analyses were made using the $Chandra$ 
Interactive Analysis of Observations (CIAO) software version 2.2.1.
Using the Level 2 processed events provided by the pipeline
processing at the $Chandra$ X-ray Center,
we selected  $ASCA$ grades 0, 2, 3, 4, and 6, as the X-ray events.
High energy electrons due to charged particles, and 
hot and flickering pixels were removed.
The effective exposure was $\sim 68$~ks for the observation.
In this paper, we concentrated on the data of S3 (BI chip) 
because this chip has the best efficiency in soft X-rays  
required for the spectral analyses,
and its on-axis position provides the best point-spread function required
for the spatial analysis.

\section*{RESULTS}

\subsection*{Image}
Figure~1 shows the images
for the NE shell of SN~1006.
The images are contrasted
in the 0.5--2.0~keV band (left) and
in the 2.0--10.0~keV band (right)
and binned to a resolution of 1~arcsec.
The fine spatial resolution of $Chandra$ unveils  
extremely narrow filaments in the hard band.   
They are running from north to south along the outer edge of the NE shell, 
parallel to the shock fronts observed
by H$\alpha$ emission line (Winkler and Long, 1997).
The soft band image, on the other hand,  
has a larger scale length
similar to the $ROSAT$ HRI image (Winkler and Long, 1997).
Many clumpy sub-structures are also seen in this energy band.

\begin{figure}
\begin{center}
\includegraphics[width=80mm]{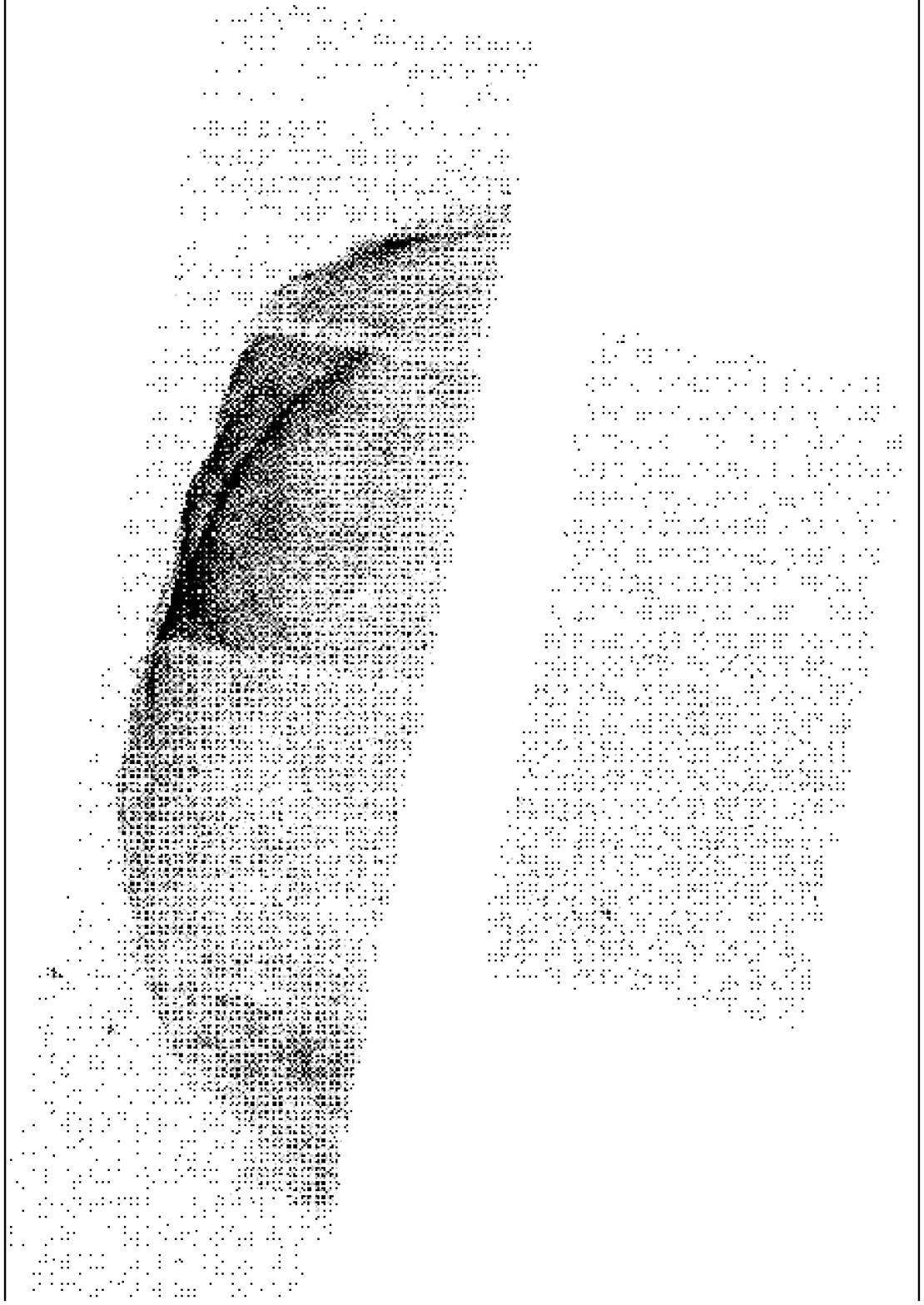}
\includegraphics[width=80mm]{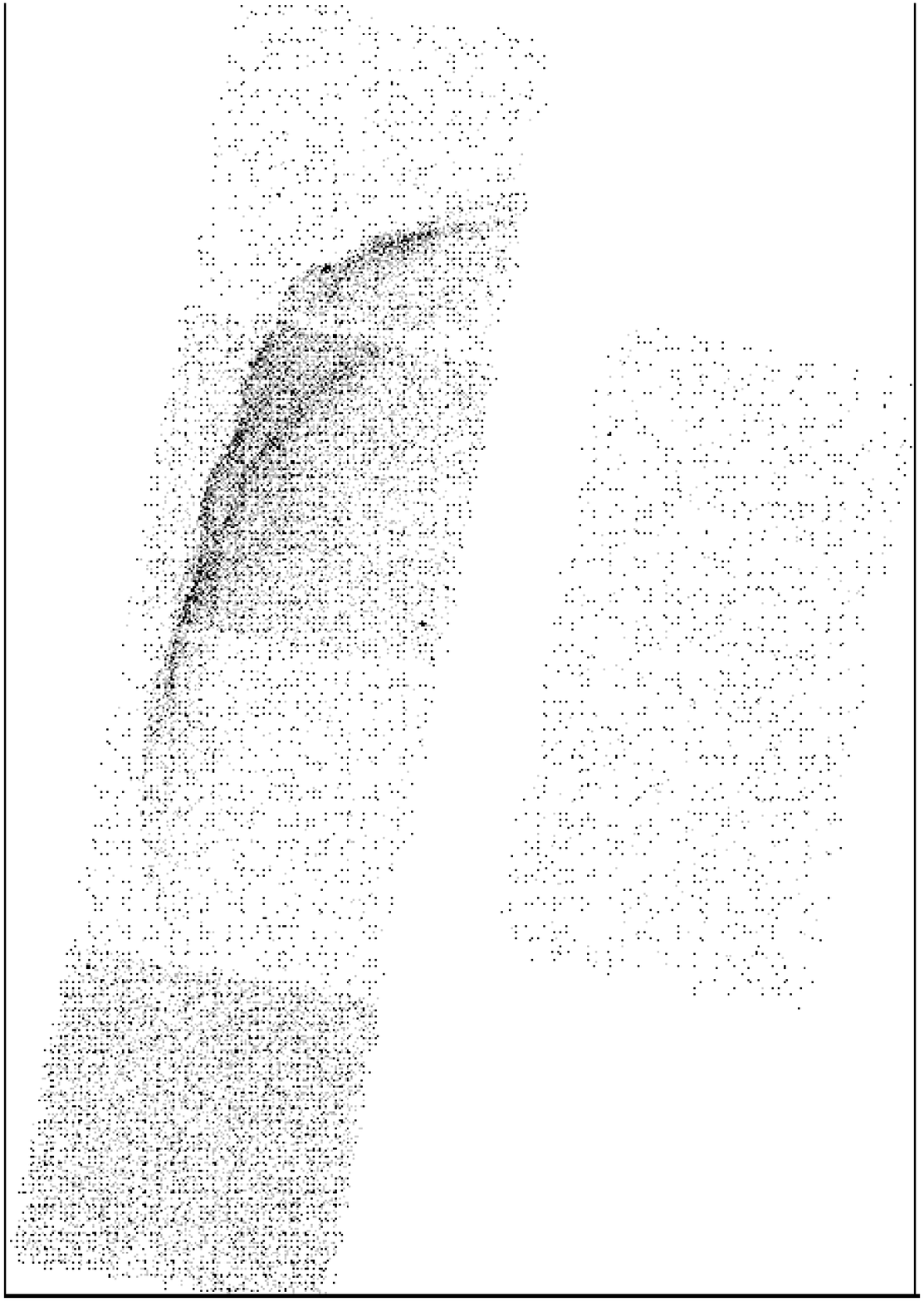}
\caption{\ \ The images of the North Eastern shell of SN~1006
binned with 1~arcsec scale,
in the 0.5-2.0~keV band (left) and the 2.0--10.0~keV band (right),
respectively,
both in logarithmic scale.}
\end{center}
\end{figure}

\subsection*{Inner Region}

To resolve the thermal and non-thermal components,
we made a spectrum from a bright clump found in the soft band image,
which is located in the inner part of the 
NE shell ( ``Inner region'' with the dashed ellipse in Figure~2).
The background region was selected from a region out of the SNR,
as is shown in Figure~2 with the dashed lines.

The background-subtracted spectrum
can be fitted with a thin thermal plasma model
in non-equilibrium ionization (NEI) state
calculated by Borkowski et al. (2001)
plus a power-law component.
The spectrum of the inner region clump is softer
than any other regions in the NE shell,
which indicates that the contribution of the thermal component is the largest.
Nevertheless the thermal photons are only 0.02\% of the non-thermal ones
if we limit the energy band to 2.0--10.0~keV (the hard band).
Therefore, in the following spatial analyses,
we regard that all the photons in the hard band
are non-thermal origin.

\subsection*{The Filaments}

The outer edge of the NE shell is outlined by several thin X-ray filaments.
For the study of these filaments, we selected 6 rectangular regions
in Figure~2,
in which the filaments are straight and free from other structures
like superimposed filament and/or clumps. 
These regions (solid boxes) are shown in Figure~2 right
with the designations of No.1--6 from north to south.
Since the SNR shell is moving (expanding) from the right to the left,
we name the right and left hand sides 
of the shock downstream and upstream
following the standard shock physics terminology.

\begin{figure}
\begin{center}
\includegraphics[width=28pc]{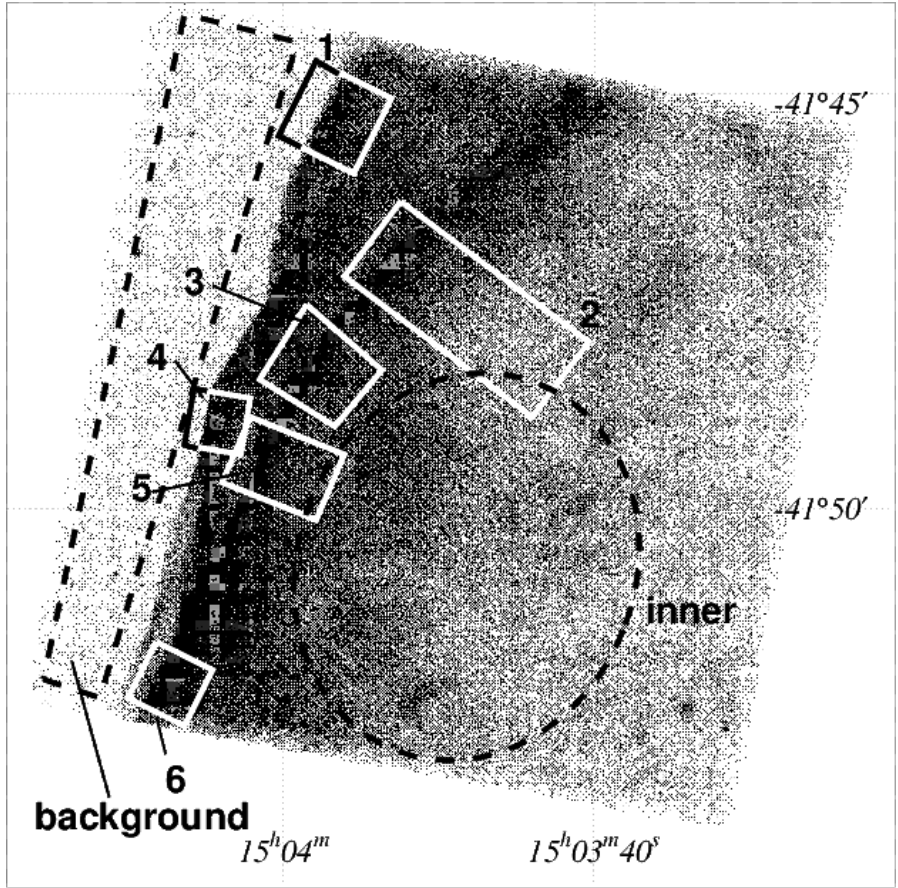}
\caption{\ \ The close-up view of the 0.5--10.0~keV band image.
The gray scale (the left bar) is given
logarithmically.
The inner and background regions for the spectral analyses
and the filament regions for the spatial analyses (No.1--6) are shown 
with dashed and solid lines, respectively.}
\end{center}
\end{figure}

Figure~3 shows the intensity profiles
in the hard (2.0--10.0~keV)
with the spatial resolution of 0.5~arcsec,
where the horizontal axis ($x$-coordinate) runs from the east to west
(upstream to downstream) along the line normal to the filaments. 
To estimate the scale length, we define a simple empirical model
as a function of position ($x$) for the profiles;
\begin{equation}
f(x) = \left\{
        \begin{array}{rlr}
        A\exp |\frac{x_0-x}{w_{\rm u}}| & {\rm in\ upstream} \\
        A\exp |\frac{x_0-x}{w_{\rm d}}| & {\rm in\ downstream},
        \end{array}
\right.
\end{equation}
where $A$ and $x_0$ are the flux and position at the filament peak.
We ignored the point spread function (PSF) of {\it Chandra} for simplicity.
The scale lengths are given by $w_{\rm u}$ and $w_{\rm d}$
for upstream and downstream, respectively.
The best-fit models are shown in Figure~3 with the solid lines.
The mean and minimum values are 0.05 and 0.01~pc for $w_{\rm u}$
and 0.19 and 0.06~pc for $w_{\rm d}$, respectively.

\begin{figure}
\begin{center}
    \includegraphics[width=0.33\textwidth]{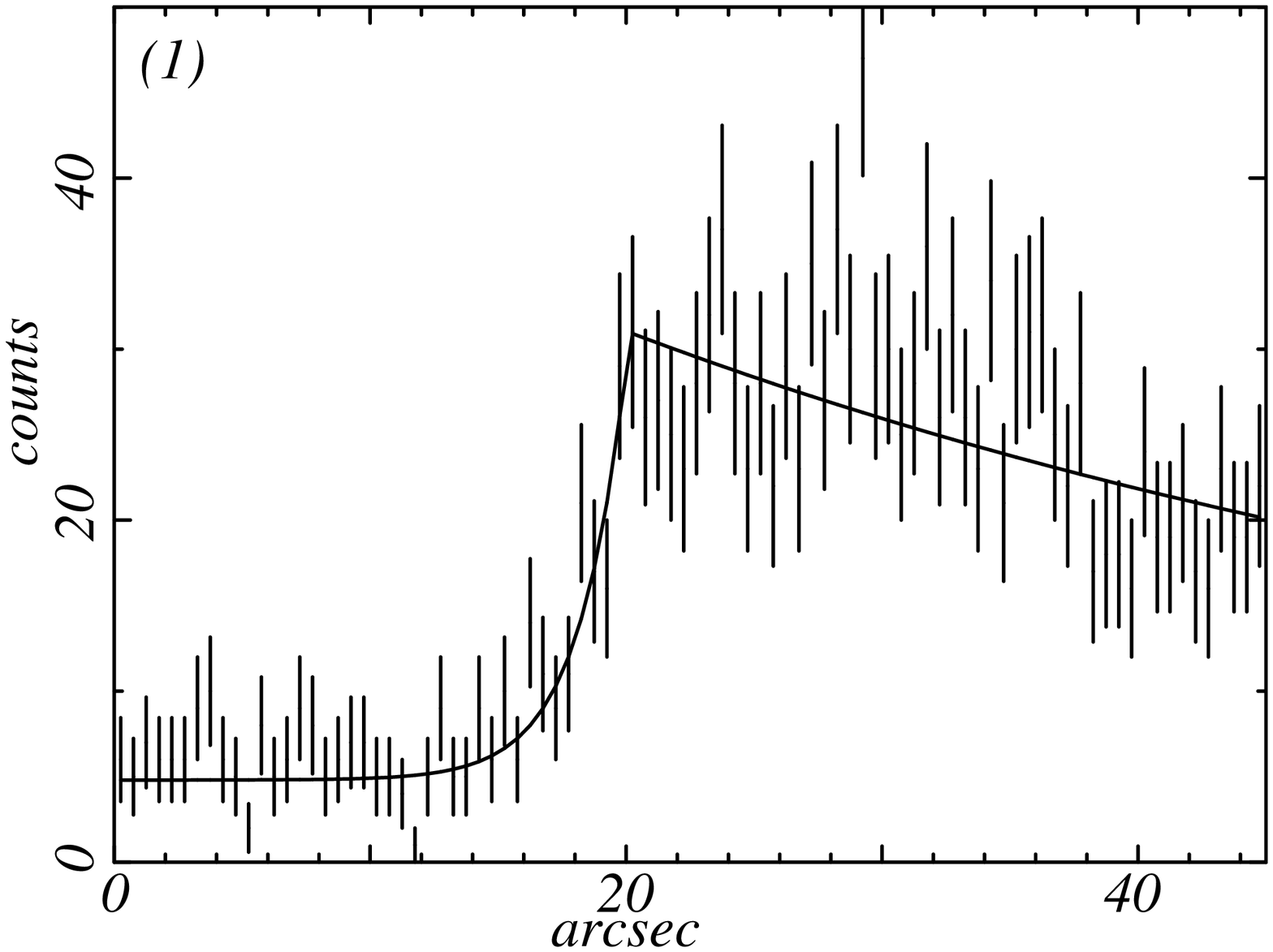}
    \includegraphics[width=0.33\textwidth]{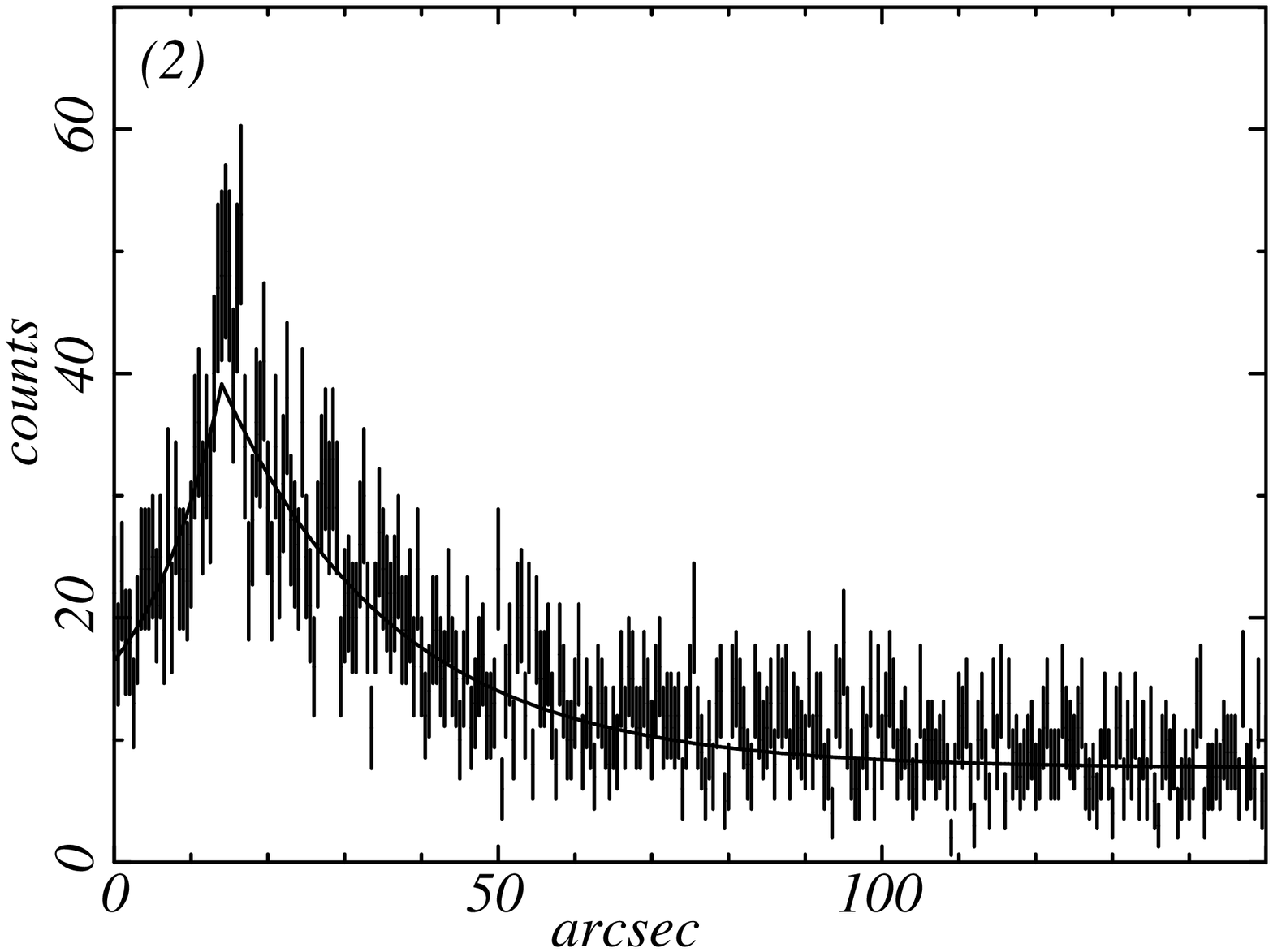}
    \includegraphics[width=0.33\textwidth]{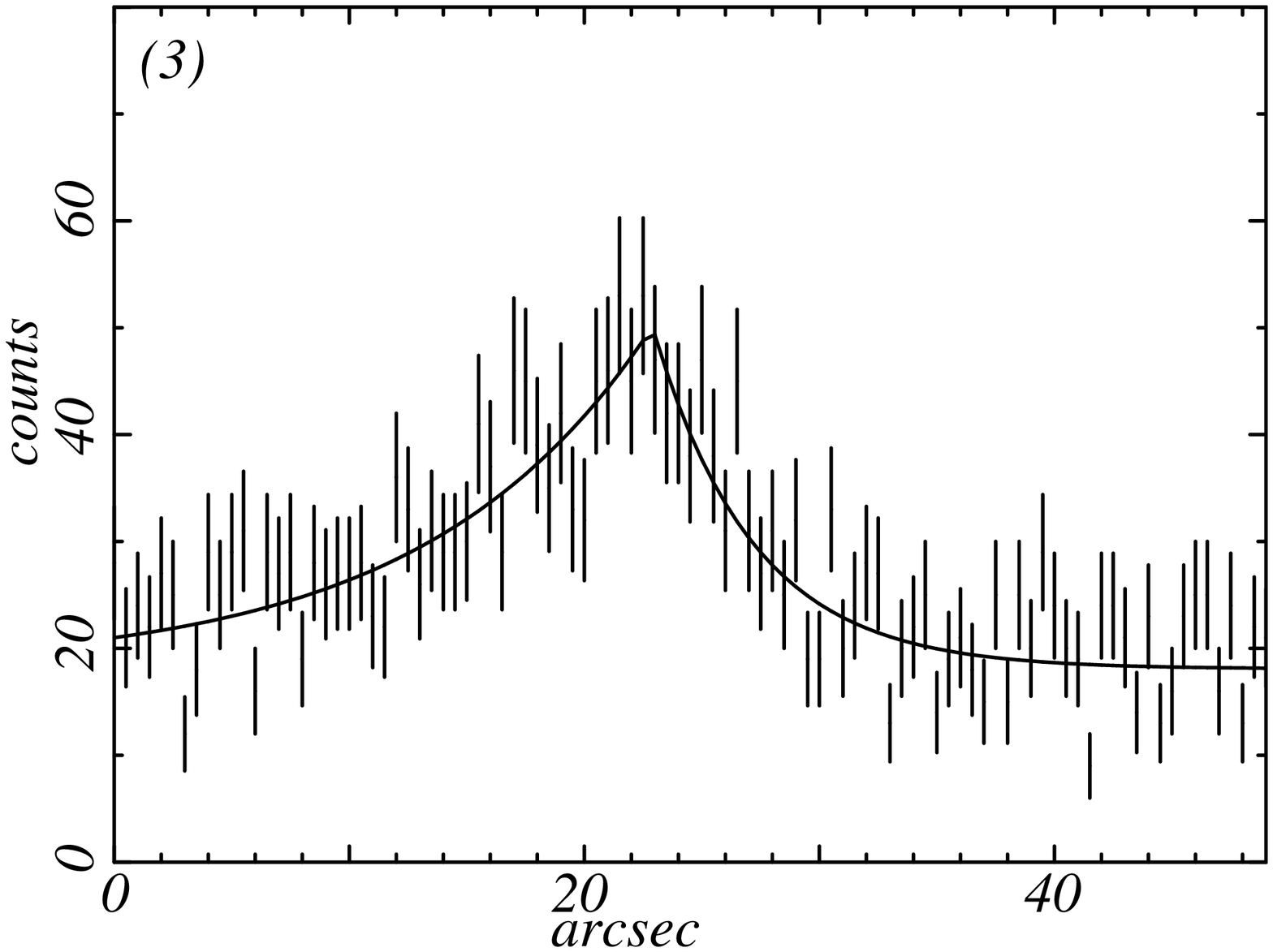}\\
    \includegraphics[width=0.33\textwidth]{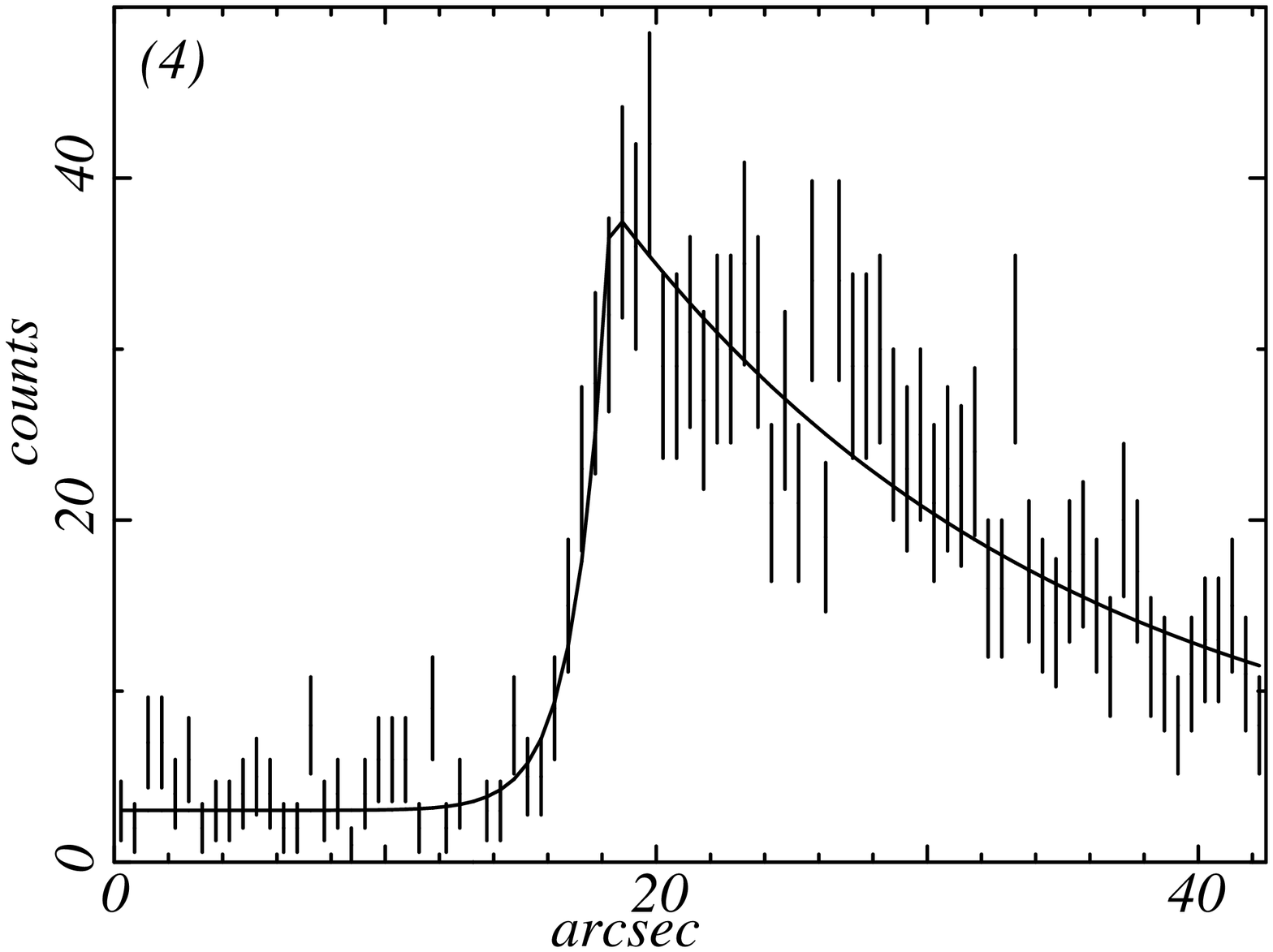}
    \includegraphics[width=0.33\textwidth]{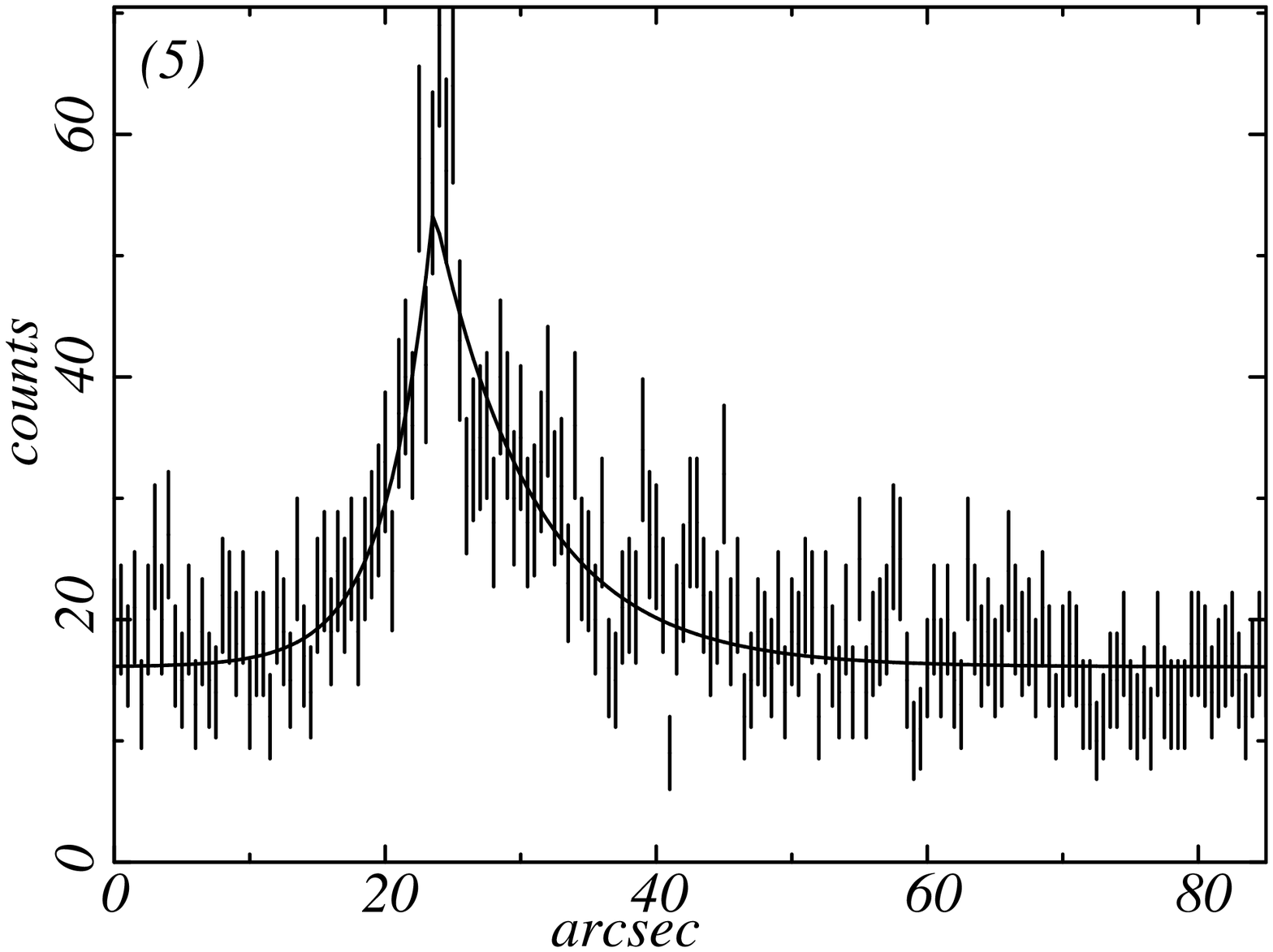}
    \includegraphics[width=0.33\textwidth]{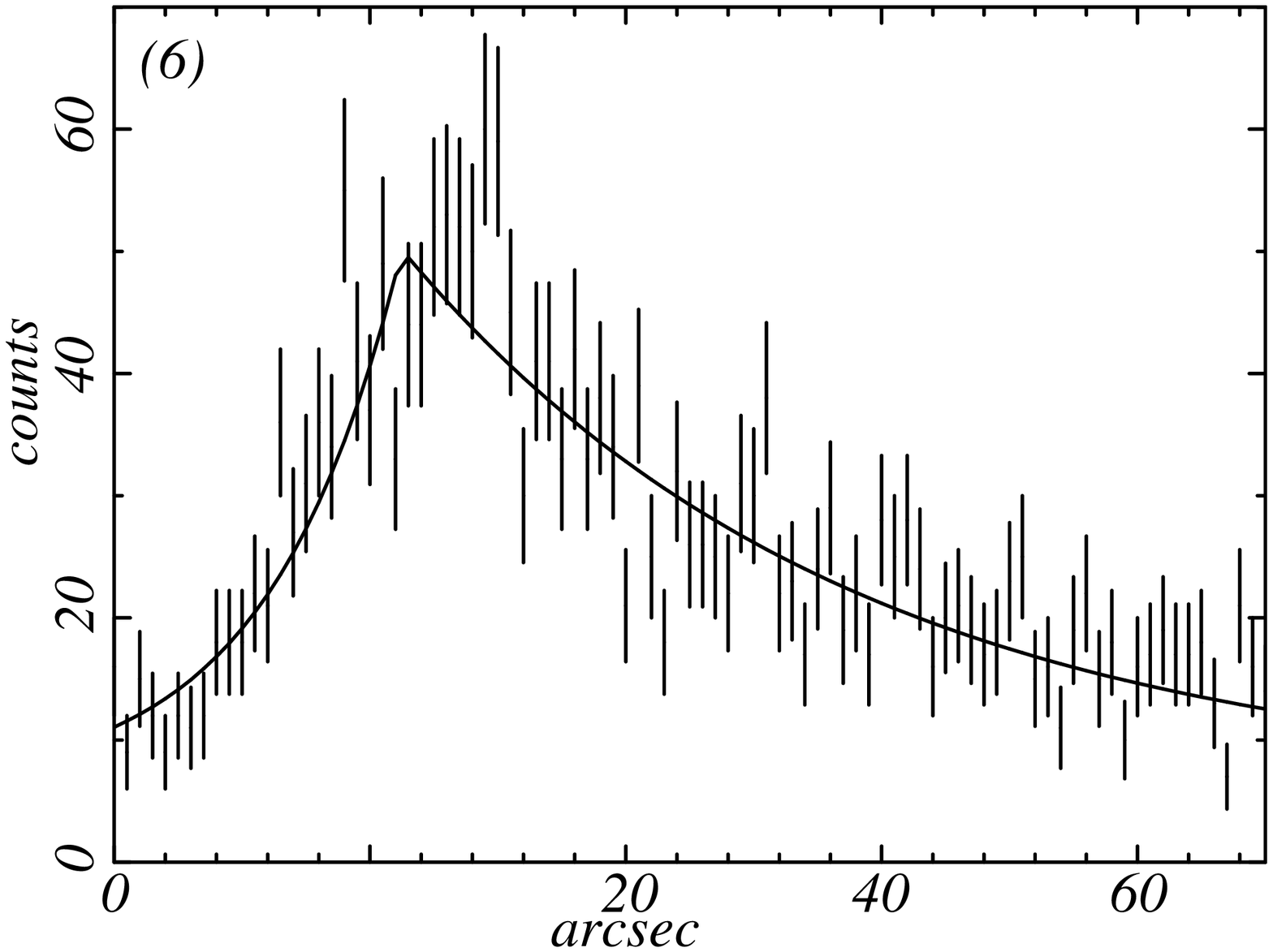}
  \caption{\ \ The profiles of the filaments in the North Eastern shell
of SN~1006 in the 2.0--10.0~keV band.}
\end{center}
\end{figure}

\section*{DISCUSSION}

In this section, we interpret the scale length of hard band X-rays,
$w_{\rm u}$ and $w_{\rm d}$.
Blandford and Orstriker (1978)
derived the distribution of accelerated particles
to be flat in downstream for a steady state.
However, we should consider the finite-time or energy loss effect
in the SN~1006 case,
where the accelerating particles must concentrate on around the shock.
Therefore, we assume that the spatial distribution
of non-thermal X-rays coincides with that of accelerated electrons.
We consider so narrower regions than the background thermal plasma,
then the magnetic field gradients can be ignored
since it may be traced by the thermal emission shown in Figure~1.
Here, we investigate the observed profiles based on 
a simple picture of DSA with the magnetic field parallel to the shock normal
and a compression ratio of about 4, and estimate the physical quantities
such as diffusion coefficient, magnetic field, and the maximum energy of
the accelerated electrons.
The scale lengths in the upstream side are
the projected values of the possible sheet-like structure,
hence real lengths should be smaller.
Considering the PSF of {\it Chandra} also makes them smaller.
We therefore adopted the minimum value of 0.01~pc.
For the scale lengths in the downstream side,
we used the mean value of 0.2~pc. 

We assume that electrons emitting synchrotron X-rays are
still accelerating on the shock front
since the previous works suggest the magnetic field is rather small
(e.g. Tanimori et al. 1998)
then the effect of synchrotron radiation may be small.
In this condition,
the diffusion coefficients in upstream ($K_{\rm u}$) and
in downstream ($K_{\rm d}$) are estimated from the relation
$w = K/u$ as following;
\begin{eqnarray}
K_{\rm u} &\simeq& w_{\rm u}\cdot u_{\rm u} = w_{\rm u}\cdot u_{
\rm s} = 8.9\times 10^{24}\ \ \ {\rm [cm^2s^{-1}]},\\
K_{\rm d} &\simeq& w_{\rm d}\cdot u_{\rm d} = w_{\rm d}\cdot
\frac{1}{4}u_{\rm s} = 4.2\times 10^{25}\ \ \ {\rm [cm^2s^{-1}]},
\end{eqnarray}
where the shock speed $u_{\rm s}$ is assumed to be 2890~km~s$^{-1}$
(Winkler et al. 2003).
The acceleration time scale $\tau_{\rm acc}$ is then
$\tau_{\rm acc} = \frac{4}{u_{\rm s}^2}(K_{\rm u}+4K_{\rm d})$
(Drury 1983);
\begin{equation}
\tau_{\rm acc} \simeq 8.6\times 10^9\ {\rm [sec]} \ = 270\ {\rm [years]}
\end{equation}
The energy loss time scale ($\tau_{\rm loss}$)
via synchrotron cooling
should be longer than the acceleration time scale
of about 300 year, hence 
using Eq.~(4), we obtain;
\begin{equation}
6.3\times 10^2 B_{\rm d}^{-2}E_{\rm max}^{-1}
\geq 8.6\times 10^9 \ \ \ {\rm [sec]}
\end{equation}
A probable region in the $E_{\rm max}$--$B_{\rm d}$ space is
then limited to the lower side of the dotted line in Figure~4.

The wide band spectra from X-ray to radio can be fitted
with the $srcut$ model (Reynolds, 1998).
The best-fit $\nu_{\rm rolloff}$ at the filaments is
2.6 (1.9--3.3)$\times 10^{17}$~Hz,
which constrains the maximum energy of electrons $E_{\rm max}$
and downstream magnetic field $B_{\rm d}$;
\begin{equation}
E_{\rm max}B_{\rm d}^{0.5} = 0.37_{-0.06}^{+0.04}\ \ \  {\rm [ergs\ G^{0.5}]}
\end{equation}
The allowed region in the $E_{\rm max}$--$B_{\rm d}$ plane 
is between the two solid lines of Figure~4.

The downstream diffusion coefficient is given by Skilling (1975) as follows;
\begin{equation}
K_{\rm d} = \frac{1}{3}\xi_{\rm d}\frac{E_{\rm max}}{{\rm e}B_{\rm d}}c,
\end{equation}
where $\xi$ is a non-dimensional parameter depending on
the angle between the shock normal and the upstream magnetic field $\theta$
and the gyro factor $\eta$,
which is the ratio of the mean free path of particles along the magnetic field
to their gyro radius (Jokipii, 1987).
Since we consider the case of $\theta=0$,
$\xi_{\rm d}$ must be $\eta_{\rm d}$ and larger than 1.
The dashed line in Figure~4
is the relation of Eq.~(7) for $\xi_{\rm d} =1$ (Bohm limit).
Therefore,
a probable region in the $E_{\rm max}$--$B_{\rm d}$ space is
to the upper of the dashed line. 

Using Eq.~(5), (6), and (7), 
we thus constrain the most likely region in the 
$E_{\rm max}$--$B_{\rm d}$ space to be in between the thick lines,
where the magnetic field in downstream ($B_{\rm d}$)
and the maximum electron energy ($E_{\rm max}$)
are in the range of 30--40~$\mu$G and 30--40~TeV, respectively.
The small $\xi_{\rm d}$
(the likely region is $1\leq\xi_{\rm d}\leq1.6$)
indicates that the magnetic field is highly turbulent.
Figure~4 also suggests that
the synchrotron loss time scale is nearly equal to the acceleration one
and electrons begin to loss their energy.

\begin{figure}
\begin{center}
    \includegraphics[height=20pc]{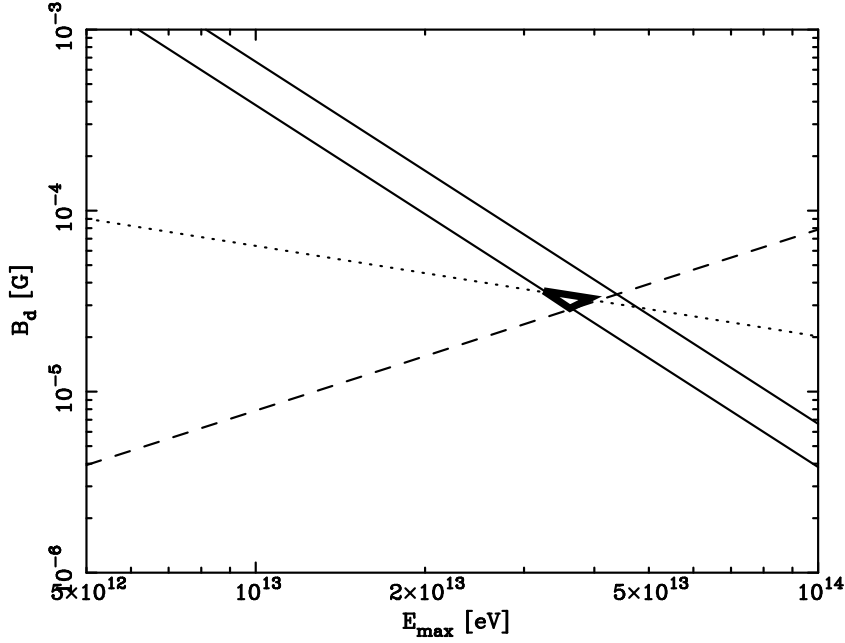}
  \caption{The relation between $E_{\rm max}$ and $B_{\rm d}$.
The solid, dotted, and dashed lines are
the relation derived from  $E_{\rm rolloff}$,
acceleration time scale,
and Bohm limit in the filaments.
The thick line region is
the most probable region for $E_{\rm max}$ and $B_{\rm d}$.}
\end{center}
\end{figure}

From Eq.~(7), the ratio of the diffusion coefficient 
($\frac{K_{\rm u}}{K_{\rm d}}$) is
given by $\frac{\xi_{\rm u}B_{\rm d}}{\xi_{\rm d}B_{\rm u}}$.
Observationally,
the ratio is about 0.2 (from Eq.~(2) and (3)),
we thus obtain the relation of magnetic field and $\xi$
between upstream and downstream as
\begin{equation}
\frac{\xi_{\rm u}}{\xi_{\rm d}} \sim  0.2\frac{B_{\rm u}}{B_{\rm d}}
\end{equation}
Since $\frac{B_{\rm u}}{B_{\rm d}} = 1$ with $\theta=0$,
and considering the large errors and many assumptions,
we find that 
$\xi_{\rm u} = \eta_{\rm u}$ may be marginally acceptable only near to 1,
while it might exist that
other possible scenarios explaining observed thin filaments.
For example, the ``apparent'' diffusion coefficients in upstream
can be extremely small
in the case of a nearly perpendicular magnetic field 
in this region as suggested by Jokipii (1987).

In this paper,
we assumed the age-limited case.
On the other hand, Figure~4 shows that the cooling process might be important
and that we might have to consider the cooling-limited case.
In such a case, protons might be accelerated to higher energy
than electrons.
This case is under investigation (Yamazaki et al. 2003).

Our constraint is more strict than determined from radio data
by Achterberg et al. (1994),
because the spatial resolution of {\it Chandra} is better than their data
and the diffusion length determined by X-rays must be larger than 
that determined by radio.
Since the current radio polarization data are spatially poor and 
limited mainly to the downstream region, fine structure observations,
particularly at the very narrow region in the upstream side should be crucial. 
Of course our result is derived in a very indirect way
with the many assumptions,
and as a result has large uncertainty,
then the DSA process might be able to explain marginally the observed results
even in parallel magnetic field.
However, we point out that another acceleration mechanism
may be able to explain our results, the thin non-thermal filaments. 
For example,
Hoshino and Shimada (2002) proposed the electron shock surfing 
acceleration mechanism.
More quantitative study is necessary,
whether the mechanism can accelerate particles up to 30~TeV in this system
and whether the accelerated particles have the power-law spectrum.

\section*{ACKNOWLEDGEMENTS}

Our particular thanks are due to
M. Hoshino, T. Terasawa, T. Yoshida, and S. Inutsuka
for their fruitful discussions and comments.
We also thank the anonymous referee
for helpful comments.
A.B., M.U., and R.Y.
are supported by JSPS Research Fellowship for Young Scientists.

\vspace{3mm}
\noindent
E-mail address of A. Bamba: bamba@cr.scphys.kyoto-u.ac.jp 

\end{document}